\documentclass[twocolumn,amsmath,amssymb,superscriptaddress,nofootinbib]{revtex4-1}
\pdfoutput=1

\usepackage[english]{babel}
\usepackage{amsthm}
\usepackage{amssymb}
\usepackage{amsmath}
\usepackage{amsthm}
\usepackage{ulem}
\usepackage[]{graphicx}
\usepackage{tensor}
\usepackage{color}
\usepackage{framed}
\usepackage{framed}
\usepackage[bookmarks,linktocpage, colorlinks=true, plainpages = false, citecolor = blue,  linkcolor=blue, urlcolor = blue, filecolor = blue]{hyperref} 
\usepackage{natbib}
\usepackage{dcolumn}
\usepackage{enumerate}
\usepackage{bm}
\usepackage{graphicx}
\usepackage{caption}
\usepackage{subcaption}
\graphicspath{{figures/}}

\theoremstyle{definition}

\newcommand{\sech}{\text{sech}}

\usepackage{tikz}
\usetikzlibrary{decorations.pathmorphing, patterns,shapes}

\input epsf

\begin{document}
\allowdisplaybreaks
\begin{titlepage}
\title{Crossing singularities in the saddle point approximation}
\author{Job Feldbrugge}
\email{job.feldbrugge@ed.ac.uk}
\affiliation{Higgs Centre for Theoretical Physics, James Clerk Maxwell Building, Edinburgh EH9 3FD, UK}
\author{Dylan L. Jow}
\email{d.jow@mail.utoronto.ca}
\affiliation{Canadian Institute for Theoretical Astrophysics, University of Toronto, 60 St. George Street, Toronto, ON M5S 3H8, Canada}
\affiliation{Department of Physics, University of Toronto, 60 St. George Street, Toronto, ON M5S 1A7, Canada}
\affiliation{Dunlap Institute for Astronomy \& Astrophysics, University of Toronto, AB 120-50 St. George Street, Toronto, ON M5S 3H4, Canada}
\author{Ue-Li Pen $^{2,\, 3,\,4,\,}$}
\email{pen@asiaa.sinica.edu.tw}
\affiliation{Institute of Astronomy and Astrophysics, Academia Sinica, Astronomy-Mathematics Building, No. 1, Section 4, Roosevelt Road, Taipei 10617, Taiwan}
\affiliation{Perimeter Institute for Theoretical Physics, 31 Caroline St. North, Waterloo, ON, Canada N2L 2Y5}
\affiliation{Canadian Institute for Advanced Research, CIFAR program in Gravitation and Cosmology}
\begin{abstract}
We describe a new phenomenon in the study of the real-time path integral, where complex classical paths hit singularities of the potential and need to be analytically continued beyond the space for which they solve the boundary value problem. We show that the behavior is universal and central to the problem of quantum tunneling. These analytically continued complex classical paths enrich the study of real-time Feynman path integrals.
\end{abstract}
\maketitle
\end{titlepage}


\textbf{Introduction:}
In the sum over histories formulation of quantum physics, evolution arises through the constructive interference of histories around classical paths. Classically allowed transitions are governed by real classical paths. Classically forbidden transitions -- such as quantum tunneling -- are often associated with complex solutions of the equations of motion interpolating between boundary conditions. Indeed, complex classical paths -- sometimes known as instantons --  are frequently studied in quantum mechanics, quantum field theory, and theories of quantum gravity. However, it is often an open question whether a complex classical solution is relevant to the real-time Feynman path integral and contributes to the corresponding amplitude. As we show in this letter, the situation is even more intricate, as the main contribution to the path integral may come from analytically continued complex classical paths that do not solve the classical boundary value problem. In this letter, we demonstrate this phenomenon for several quantum systems and discuss its various implications.

We consider tunneling of a wave packet through a barrier centered at $x=0$.  This can be considered an initial value problem of the time-dependent Schr\"odinger equation, with an initial Gaussian, minimum-uncertainty wave packet on the left with positive momentum.  For high-momentum wave packets, most of the probability passes over the barrier, similar to the classical equation of motion.  For small momenta, with energies less than the height of the barrier, most of the wave packet reflects back, and a small amount tunnels. 

The time-dependent Schr\"odinger equation,
\begin{align}
    i \hbar\frac{\partial\psi_t(x)}{\partial t}=\hat{H} \psi_t(x),
\end{align}
is solved by the propagator
\begin{align}
    \psi_T(x_1)&=\int \psi_0(x_0) G[x_0,x_1;T] \mathrm{d} x_0\,,\\
    G[x_0,x_1;T]
    &=\int_{x(0)=x_0}^{x(T)=x_1} e^{i S[x]/\hbar}\mathcal{D}x\,.
    \label{eqn:path}
\end{align}
The second equality (\ref{eqn:path}) is the Feynman path integral formulation, which for small $\hbar$ is solved by stationary points of the action $\delta S/\delta x=0$ resulting in Euler-Lagrange (EL) equations of motion 
\begin{align}
    m \ddot{x}=-V'(x)\,,
    \label{eqn:EL}
\end{align}
with the mass $m$, the Heaviside theta function $\Theta_H$ and the dots denoting time derivatives, satisfying boundary conditions $x(0)=x_0,\ x(T)=x_1$ \cite{Feynman:1948,Feynman:1965,Feynman:1948b}. In the Picard-Lefshetz (PL) picture, one deforms the real space path integral (\ref{eqn:path}) contour into the sum of complex path segments (thimbles), each of which is non-oscillatory \cite{Turok:2014,Feldbrugge:2023}.  For small $\hbar$, each thimble is dominated by its saddle point. This saddle may be complex, even though the original integral was defined over real paths.  It is tempting to consider complex solutions (saddles) of (\ref{eqn:EL}) satisfying the real boundary conditions for complex energies and evaluate the propagator (\ref{eqn:path}) using the corresponding action \cite{Schulman:2012}.  Two technical complications arise: 1. while PL saddles are solutions to the EL equations, the converse is generally not true, and 2., no constructive example has been found where the propagator from the Schr\"odinger equation agrees with the complex saddle. 

In this paper we revisit this problem, systematically examining the real and complex paths for fixed $x_0$ while varying $x_1$. The Rosen-Morse barrier (see below) has a particularly simple exact solution (saddle) set, and corresponding exact actions at those saddles. This allows us to examine the propagator under these continuous deformations. 

Several conceptual issues arise: 1. complex saddles depend on variable rescalings, and 2. the evaluation of the action at a complex saddle gives a different value from the analytic continuation of the action along the deformation path of (real) $x_1$.  We find these are reconciled due to branch cuts and singularity crossings for the saddles.

This Letter addresses this novel approach of treating branch cuts in the action, explicated by the exact Rosen-Morse barrier solution, but generalizable to generic potentials.  The analysis is simplified using the propagator in Energy $E$, conjugate to time $T$.

\textbf{The Rosen-Morse barrier:}
The symmetric Rosen-Morse \cite{Rosen:1932} or modified P\"oschl-Teller system \cite{Poschl:1933}, describing the evolution of a non-relativistic particle in the potential 
\begin{align}
    V(x)=V_0\, \sech^2(x)\,,
\end{align}
is one of the few fully solvable non-linear quantum systems. For simplicity, we will restrict ourselves to the barrier problem with $V_0 \geq \hbar^2/(8m)$. The real-time energy propagator of a non-relativistic particle with mass $m$ propagating form $x_0$ to $x_1$ at energy $E$ is solved in closed form \cite{Kleinert:1992},
\begin{align}
    K[x_1,x_0;E] &= \int_0^\infty \int_{x(0)=x_0}^{x(1)=x_1} e^{i (S[x] + E T ) /\hbar}\mathcal{D}x\, \mathrm{d}T\\
    &= -\frac{im \Gamma(-i k_E - N)\Gamma(- i k_E + N +1)}{\hbar} \nonumber\\
    &\phantom{=} \times P_{N}^{i k_E}(\tanh x_>)P_{N}^{i k_E}(-\tanh x_<)\,,
\end{align}
where the path integral ranges over the continuous paths interpolating between $x(0)=x_0$ and $x(1)=x_1$ parametrized by $\lambda \in [0,1]$, and the action
\begin{align}
    S[x] = \int_0^1 \left[\frac{m x'(\lambda)^2}{2T} - T V(x(\lambda)) \right]\mathrm{d}\lambda\,,
\end{align}
with the propagation time $T$, and the prime denoting derivatives with respect to the parameter $\lambda$. The closed-form solution includes the associated Legendre function $P_\lambda^\mu(x)$ with the parameter $N=-\tfrac{1}{2} + \tfrac{i}{2\hbar} \sqrt{8 m V_0-\hbar^2}\,,$ the minimum $x_<$ and the maximum $x_>$ of the boundary conditions $(x_0,x_1)$, and dimensionless momentum $k_E= \sqrt{2m E}/\hbar$. The Legendre function with degree $\text{Re}[\lambda]=-1/2$ is known as the conical or Mehler function. 

\begin{figure*}
    \centering
    \begin{subfigure}[t]{0.45\linewidth}
        \includegraphics[width =\textwidth]{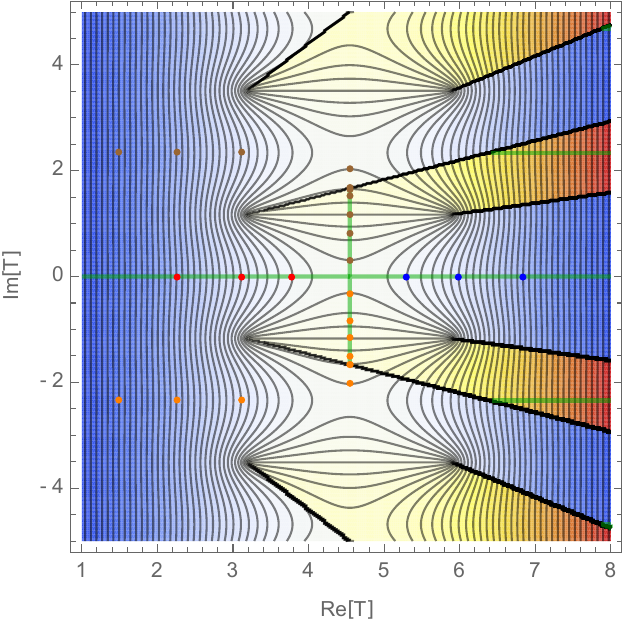}
        \caption{$x_1$}
    \end{subfigure}~
    \begin{subfigure}[t]{0.45\linewidth}
        \includegraphics[width =\textwidth]{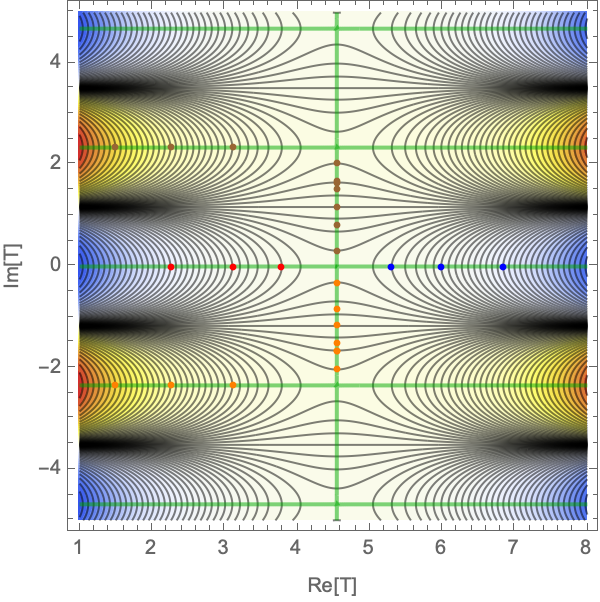}
        \caption{$\sinh x_1$}
    \end{subfigure}
    \caption{The final position of the classical paths starting at $x_0=-5$ for an energy $E=0.9$, mass $m=1$, and a barrier strength $V_0=1$ in the complex $T$ plane. The green lines indicate times that result into a real final position.}\label{fig:Complex_T}
    \begin{subfigure}[t]{0.45 \linewidth}
        \includegraphics[width=\linewidth]{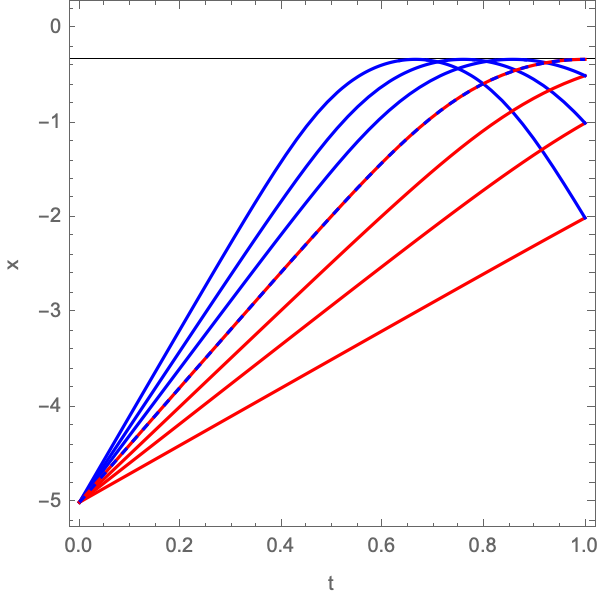}
        \caption{Real classical paths}\label{fig:realPathsRosen}
    \end{subfigure}~
    \begin{subfigure}[t]{0.45 \linewidth}
        \includegraphics[width=\linewidth]{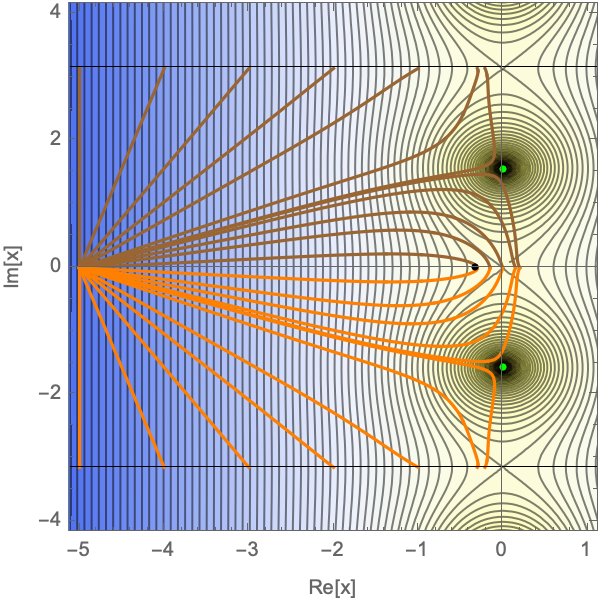}
        \caption{Complex classical paths}\label{fig:complexPathsRosen}
    \end{subfigure}
    \caption{The classical paths of the energy propagator at $E=0.9$ for a particle starting at $x_0=-5$ with mass $m=1$ in a barrier with strength $V_0=1$. \textit{Left:} the real classical paths for the final positions $x_1 = -2, -1, -0.5,-x_c$. \text{Right:} the complex classical paths for the final positions $x_1= -0.3, -0.15, 0, 0.15, 0.2$ and the analytically continued paths for $x_1=0.21, 0.3, 1, 2, 3, 4, 5$. The orange and brown paths are conjugate pairs. The black dot represents the turning point, the background is the absolute valued squared of the potential and the green points are the singularities of the potential.}
    \label{fig:ClassicalPathsRosen}
\end{figure*}

The associated classical system is given by the Euler-Lagrange and Hamilton-Jacobi equations
\begin{align}
    \frac{\delta S}{\delta x} = 0\,,\quad E+ \frac{\partial S}{\partial T}=0\,,
\end{align}
which for the Rosen-Morse system assumes the form
\begin{align}
    \frac{m x''}{T^2} = \frac{2 V_0 \tanh x}{\cosh^2 x}\,, \quad
    E =\frac{m x'(0)^2}{2T^2} + \frac{V_0}{\cosh^2 x_0}\,.
\end{align}
The Euler-Lagrange equation is solved by
\begin{align}
    \sinh x_C(\lambda) = c \sqrt{\frac{E-V_0}{E}} \sinh\left[ \sqrt{\frac{2E}{m}}(T \lambda -C) \right]\,,
\end{align}
with the sign $c = \pm 1$ and the shift parameter $C$, with the associated classical action 
\begin{align}
    &S_C =  ET -\sqrt{2m V_0} \nonumber \\
    &\times \tanh^{-1}\left[\sqrt{\frac{V_0}{E}} \tanh\left[\sqrt{\frac{2 E}{m}}(T\lambda-C) \right]\right]\Bigg|_{0}^1\,.
    \label{eq:classicalAction}
\end{align}
We solve the initial value problem, where the Hamilton-Jacobi equation yields the initial velocity
\begin{align} 
    v_0= \pm \sqrt{\frac{2}{m}\left[E - V_0\, \sech^2 x(0)\right]}\, T\,,
\end{align}
for complex $T$ to find the solutions to the boundary value problem associated with the energy propagator. Candidate classical solutions lay on the curves for which the final position $x(\lambda=1)$ is real (see the green curves in the left panel of fig.\ \ref{fig:Complex_T}).

When the energy is below the top of the barrier ($E < V_0$) and the initial and final position lay in the same classically allowed region (either $x_0,x_1 \leq -x_c$ or $x_c \leq x_0, x_1$ with the turning point $x_c = \cosh^{-1} \sqrt{V_0/E}$) there exist two real classical solutions to the boundary value problem. The red and blue points in fig.\ \ref{fig:Complex_T} represent the direct and bouncing solution of the boundary value problem (see the left panel of fig.\ \ref{fig:ClassicalPathsRosen}). When keeping the initial position fixed in the left region and letting the final position $x_1$ approach the turning point $x_c$, the two real classical paths coalesce. For larger final positions, no classical solution exists, as the real paths have formed a conjugate pair of complex classical paths (see the brown and orange points in the left panel of fig.\ \ref{fig:Complex_T}). According to Picard-Lefschetz theory, the complex classical path for which the exponent of the path integral $i(S + ET)$ has a negative real part (the orange path) remains relevant after this transition. The other complex path (marked by the brown points) is irrelevant to the path integral. 

As we increase the final position further, the complex classical path moves further into the complex plane along the vertical green line, until it collides with a discontinuity in the final position in the complex $T$-plane. For larger $x_1$, the solution to the boundary value problem ceases to exist. This dramatic transition occurs as the complex classical path intersects one of the singularities of the analytic continuation of the potential at $i\pi (\tfrac{1}{2} +n)$ with $n \in \mathbb{Z}$ (see the right panel of fig.\ \ref{fig:ClassicalPathsRosen}). At this \textit{singularity crossing}, the naive saddle point approximation of the path integral \cite{Schulman:2012},
\begin{align}
    K \sim \sum_{(x_C,T_C)} \sqrt{-\frac{\partial^2 S_C/\partial x_0 \partial x_1}{\partial^2S_C/\partial T^2}} e^{i(S_C+ET_C)/\hbar}\,,
\end{align}
ranging over the relevant classical paths fails to approximate the tunneling behavior of the energy propagator. To continue the saddle point approximation through this singularity crossing, we interpret the discontinuity (see the left panel of fig.\ \ref{fig:Complex_T}) as a branch cut and analytically continue the saddle point beyond the space of classical paths satisfying the boundary conditions. For the Rosen-Morse system, this analytical continuation is most easily implemented by working with the hyperbolic sine of the final position $\sinh x_1$ (see the right panel of fig.\ \ref{fig:Complex_T}). The hyperbolic sine function unveils the analytic continuation saddle point and highlights the occurrence of a complex caustic, where the horizontal and vertical green lines intersect away from the real line, corresponding to $x_1$ moving into the classically allowed region on the right. As far as we are aware, this is the first example of a system with a complex caustic, where complex saddle points coalesce. After the complex caustic, the relevant saddle point moves to the left along the horizontal green line.

Beyond the singularity crossing, the complex $T$ only serves as a label of the analytically continued saddle point, as the corresponding initial value problem does not terminate at the boundary condition $x(1)=x_1$. Instead, the solution to the equation of motion approaches $x(1)=-x_1- i \pi$ (see the right panel of fig.\ \ref{fig:ClassicalPathsRosen}), as $\sinh(-x_1 - i \pi) = \sinh(x_1)$. Note that when approximating the energy propagator for the configuration $x_1=-x_0$ for small $x_0$, the solution to the initial value problem linearly interpolates between $x_0$ and $x_0-i\pi$. To describe the tunneling behavior, the classical action also needs to be analytically continued at the singularity crossing. In the classical action \eqref{eq:classicalAction}, the singularity crossing corresponds to a branch-cut of the arctanh function, yielding a correction $i \sqrt{2m V_0}\pi$ to the classical action of the solution of the initial value problem.

\begin{figure}
    \includegraphics[width =\linewidth]{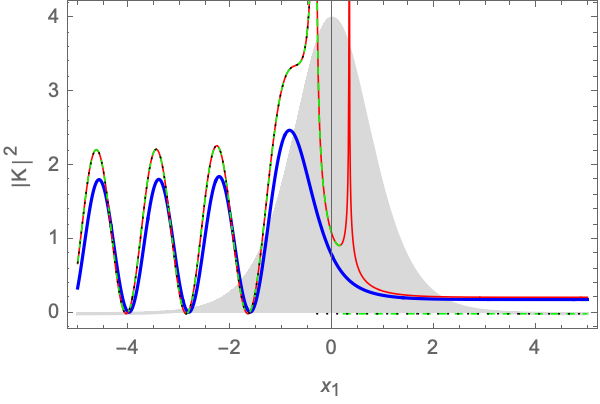}
    \caption{A comparison of the energy propagator $E=0.9$ as a funciton of $x_1$ and the saddle point approximations for the initial position $x_0=-5$, for a particle of mass $m=1$ in a barrier of strength $V_0=1$ and the reduced Planck constant $\hbar=0.5$. The energy propagator (blue), the real saddle point approximation (dotted black), the saddle point approximation with complex solutions to the boundary value problem (dashed green), and the saddle point approximation including the analytic continuations of the complex saddle points (red).}\label{fig:RosenPropagator}
\end{figure}

We compare the exact energy propagator for a particle below the top of the barrier in fig.\ \ref{fig:RosenPropagator}. In the classically allowed region, we observe an interference pattern that can be understood in terms of the direct and bouncing real classical paths. At the turning point, $x_1=x_c$, the saddle point approximation diverges in a fold caustic. For larger $x_1$ the real saddle point approximation vanishes. The complex saddles reasonably approximate the exact propagator until the singularity crossing, where the green line drops to zero. After the singularity crossing, the saddle point approximation with the analytically continued classical path diverges in a second complex caustic, when $x_1$ enters the classically allowed region after which the approximation approaches the non-zero exact propagator (this value is proportional to the tunneling amplitude). From this analysis, it is clear that the singularity crossing and the corresponding analytic continuation are central to quantum tunneling in the real-time Feynman path integral. The same can be said for quantum reflections when the energy of the particle exceeds the potential strength, $E > V_0$.

Indeed, as we show in \cite{Feldbrugge:2023Rosen}, the WKB tunneling rate is only recovered when including the correction $i \sqrt{2m V_0}\pi$ in the saddle point approximation.   Note that the saddle point approximation and the exact result do not coincide as we are working at finite $\hbar$ and with an energy close to the top of the barrier. The saddle point approximation converges to the exact propagator away from the caustics for lower energies and reduced Planck constants, however, these changes will also suppress the tunneling amplitude and energy propagator for large $x_1$.

\textbf{The Gaussian barrier:}
The path integral of the Rosen-Morse potential exhibits a singularity crossing when a relevant complex classical path intersects a singularity of the analytically continued action. Similar behavior can be observed for rational potentials, such as the Lorentzian potential $V(x)=V_0/(1+x^2)$ with singularities at $x = \pm i$. However, singularity crossings are not restricted to real-time path integrals of theories whose analytically continued potential has singularities at finite distances in the complex plane. 

\begin{figure}
    \includegraphics[width =\linewidth]{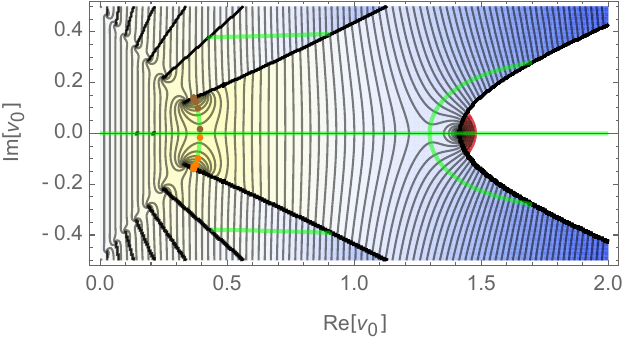}
    \caption{Classical paths for the Gaussian barrier in the complex initial velocity plane for the initial position $x_0=-5$, particle mass $m=1$, the barrier strength $V_0=1$, and the total propagation time $T=10$. The green curves indicate initial velocities that lead to a real final position.}\label{fig:Complex_v0}
\end{figure}

To illustrate this, consider a particle interacting with a Gaussian barrier $V(x)=V_0 \exp(-x^2)$ for $V_0 >0$, which has an essential singularity at complex infinity. Let's consider the real-time path integral
\begin{align}
    G[x_1,x_0;T] =  \int_{x(0)=0}^{x(1)=x_1} e^{i S[x]/\hbar}\mathcal{D}x\,,
\end{align}
with the associated Euler-Lagrange equation
\begin{align}
    \frac{\delta S}{\delta x} = 0:\quad m x'' = 2V_0 T^2  x e^{-x^2}\,,
\end{align}
with the boundary conditions $x(0)=x_0$ and $x(1)=x_1$ for fixed real propagation time $T$. We look for solutions to the boundary value problem by solving the initial value problem in the space of complex initial velocities $x'(0)=v_0$ (see fig.\ \ref{fig:Complex_v0}). 

\begin{figure}
    \includegraphics[width =\linewidth]{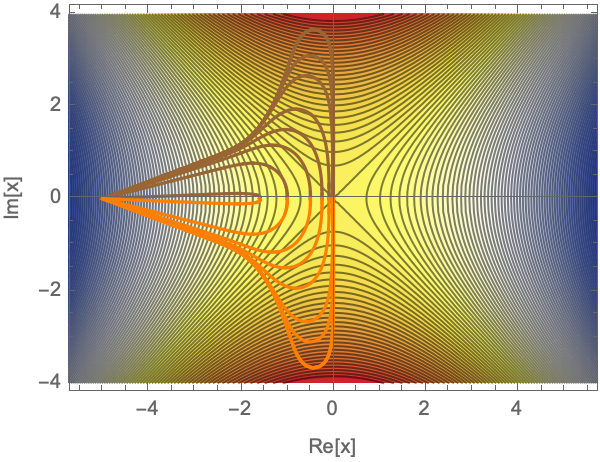}
    \caption{Complex classical paths (orange and green are conjugate paths) for the Gaussian barrier in the complex initial velocity plane for the initial position $x_0=-5$, particle mass $m=1$, the barrier strength $V_0=1$, the total propagation time $T=10$, and the final position $x_1=-1.6,$ $-1,$ $-0.5,$ $-0.25,$ $-0.1,$ $-0.03,$ $-0.02,$ $-0.01$}\label{fig:Gaussian_paths}
\end{figure}

When the initial and final positions reside on the same side of the barrier, there either exists a single or three real classical paths separated by a caustic. When there are three real classical paths, the particle can either travel directly between the boundary points, travel with a light bounce, or spend a significant amount of time near the top of the barrier before rolling back to the final position. Keeping the initial position fixed on the left of the barrier, and moving the final position to the top of the barrier, two of the three real classical paths coalesce in a caustic at the turning point. For larger $x_1$, the real classical paths form complex conjugate pairs of which one remains relevant to the Feynman path integral. As we see in fig.\ \ref{fig:Complex_v0} the complex path again hits discontinuities of the final position as a function of the complex initial velocity. As the complex path approaches the discontinuity (see fig.\ \ref{fig:Complex_v0}), the classical path travels to complex infinity and back in finite time (see the paths in fig.\ \ref{fig:Gaussian_paths}). The real-time Feynman path integral of the Gaussian barrier thus includes complex paths undergoing singularity crossings where the path runs off to complex infinity.

\textbf{The discretized path integral:} Singularity crossings of complex classical paths are universal in the continuum path integral but absent in the simplest discretized version. Consider the time-discretized path integral
\begin{align}
    \int e^{\frac{i \epsilon T}{\hbar}\sum_{j=0}^{n-1}\left[\frac{m}{2}\left(\frac{y_{j+1}-y_j}{\epsilon T}\right)^2 - V(y_j)\right] }\mathrm{d}y_1\mathrm{d}y_2\dots \mathrm{d}y_{n-1}\,,
\end{align}
with $n$ equially space timesteps $y_j = x(\epsilon j)$ for $j=0,1,\dots,n$ of size $\epsilon = 1/n$ with $y_0=x_0$ and $y_n=x_1$. This discretization resembles a multi-plane lens system in wave optics \cite{Feldbrugge:2023MNRAS.520.2995F}. The variation of the action yields the discrete equation of motion
\begin{align}
    \frac{y_{j+1}-y_j}{\epsilon T} = \frac{y_{j} - y_{j-1}}{\epsilon T} - \frac{\epsilon T}{m} V'(y_i)\,,
\end{align}
for $j=1,2,\dots,n-1$. This set of equations has a unique recursive solution for $y_2,\dots, y_n$ given the initial position $y_0$ and the discrete initial velocity $\bar{v}_0 = (y_1 - y_0)/\epsilon $ resembling the initial value problem,
\begin{align}
    y_j 
    &= x_0 + j \epsilon  \bar{v}_0 - \frac{\epsilon^2 T^2}{m} \sum_{k=1}^{j-1} (j-k) V'(y_k)\,,
\end{align}
for $j=2,\dots,n$. When the potential is analytic and free of branch cuts, such as $V(x)=\sech^2(x)$, so is the final position $y_n$ as a function of the initial position and discretized initial velocity. Consequently, the solutions of the discretized Rosen-Morse system never undergo singularity crossings, making it qualitatively different from the continuum theory.

While the discretized theory does not have singularity crossings, the final position $y_n$ does develop a large set of singularities in the complex $\bar{v}_0$ plane and a large set of additional complex solutions to the boundary value problem, rapidly increasing with the number of discretization steps. When tracking a singularity crossing event in the continuum theory, we find that the corresponding discretized path develops large jumps, invalidating the discretization condition. Further research is required to verify whether the absence of singularity crossings in the discretized theory is replaced by additional relevant complex discretized classical paths.

\textbf{Implications:}
The real-time Feynman path integral studied in terms of classical paths not only includes real and complex but also analytically continued paths. Beyond the singularity crossing, the complex classical path no longer solves the boundary value problem, yet can dominate the path integral. We believe that previous attempts to express classically forbidden phenomena with complex paths have been frustrated by the lack of this observation. When interpreting such an analytically continued path in terms of an initial value problem, the termination point is altered and additional terms need to be added to the classical action and functional determinant for the saddle point approximation to hold. 

This phenomenon has large implications in the study of instantons, particularly in quantum gravity where complex solutions of the Einstein field equations are commonly studied \cite{Coleman:1980, Hawking:1982, Hartle:1983, Vilenkin:1982,Feldbrugge:2017}. Since general relativity predicts curvature singularities, complex solutions will likely undergo singularity crossings affecting the path integral for gravity. Witten recently proposed a complex space-time selection criterion \cite{Witten:2021} based on the study of quantum fields on complex metrics \cite{Louko:1997, Kontsevich:2021} which will need to be reconsidered in light of these singularity crossings; how does one embed a quantum field on an analytically continued spacetime, when the solution to the Einstein field equations satisfying the boundary conditions does not exist?

\section*{Acknowledgements}
The work of JF is supported by the STFC Consolidated Grant ‘Particle Physics at the Higgs Centre,’ and, respectively, by a Higgs Fellowship and the Higgs Chair of Theoretical Physics at the University of Edinburgh.

For the purpose of open access, the author has applied a Creative Commons Attribution (CC BY) license to any Author Accepted Manuscript version arising from this submission.


\bibliographystyle{utphys}
\bibliography{Library}

\newpage
\section{Supplemenatary material}
\begin{figure*}
    \begin{subfigure}[t]{0.32\linewidth}
        \includegraphics[width =\textwidth]{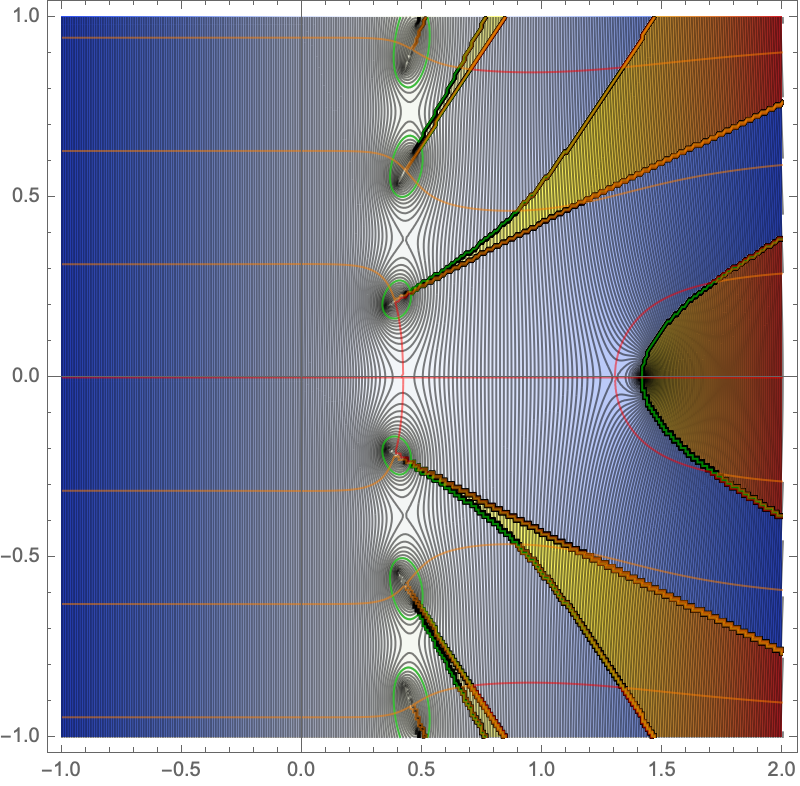}
        \caption{Continuum}\label{fig:continuumD}
    \end{subfigure}~
    \begin{subfigure}[t]{0.32\linewidth}
        \includegraphics[width =\textwidth]{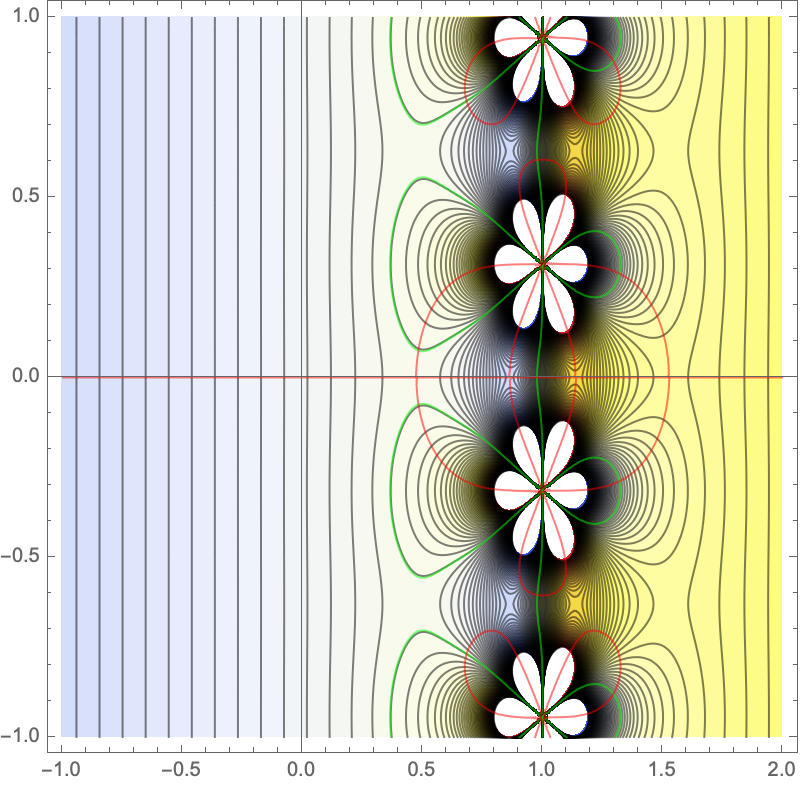}
        \caption{$n=2$}
    \end{subfigure}~
    \begin{subfigure}[t]{0.32\linewidth}
        \includegraphics[width =\textwidth]{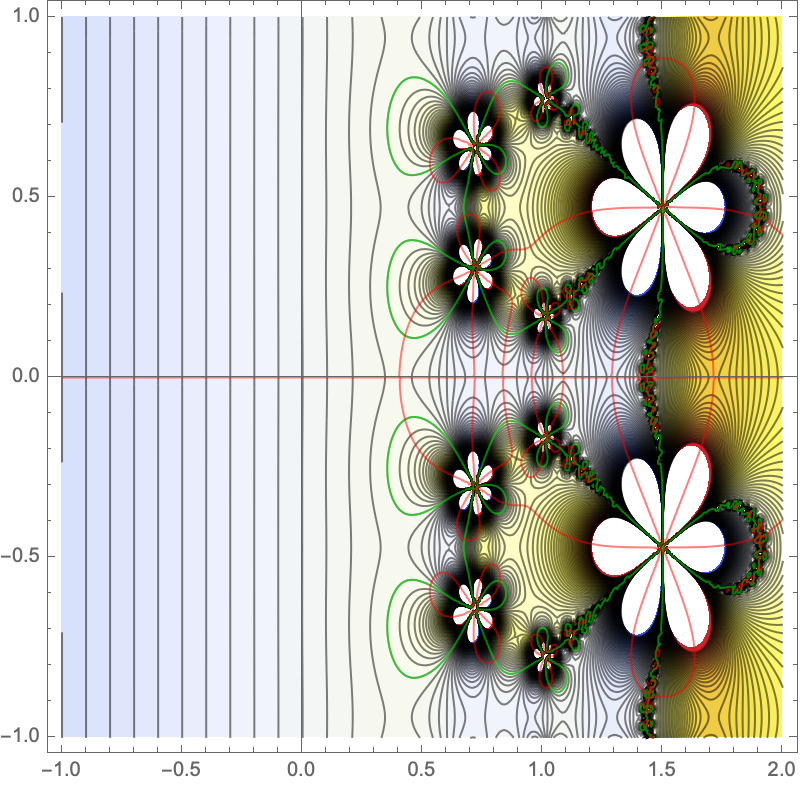}
        \caption{$n=3$}
    \end{subfigure}\\
    \begin{subfigure}[t]{0.32\linewidth}
        \includegraphics[width =\textwidth]{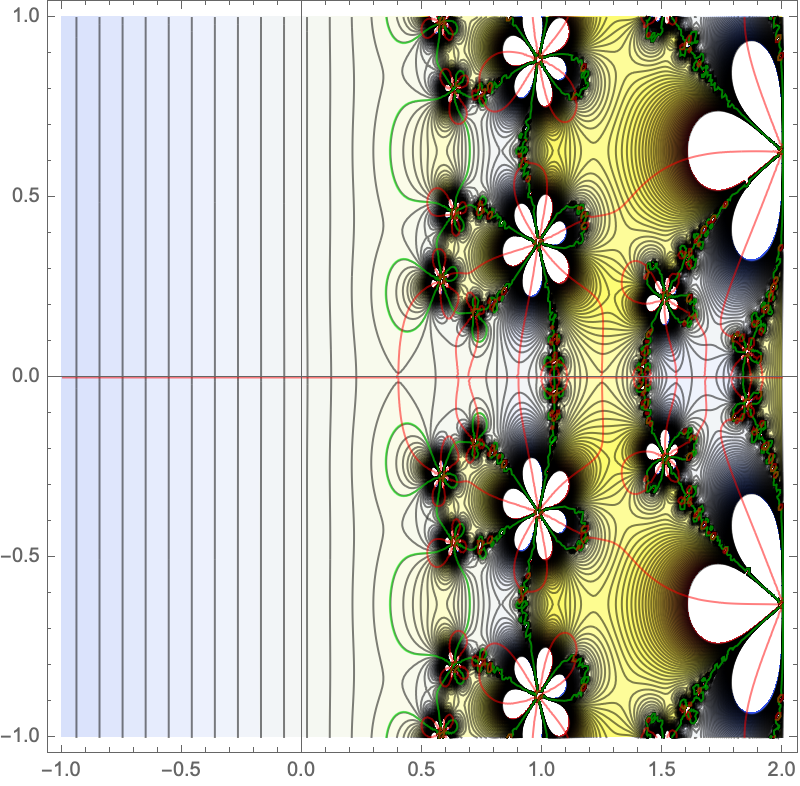}
        \caption{$n=4$}
    \end{subfigure}~
    \begin{subfigure}[t]{0.32\linewidth}
        \includegraphics[width =\textwidth]{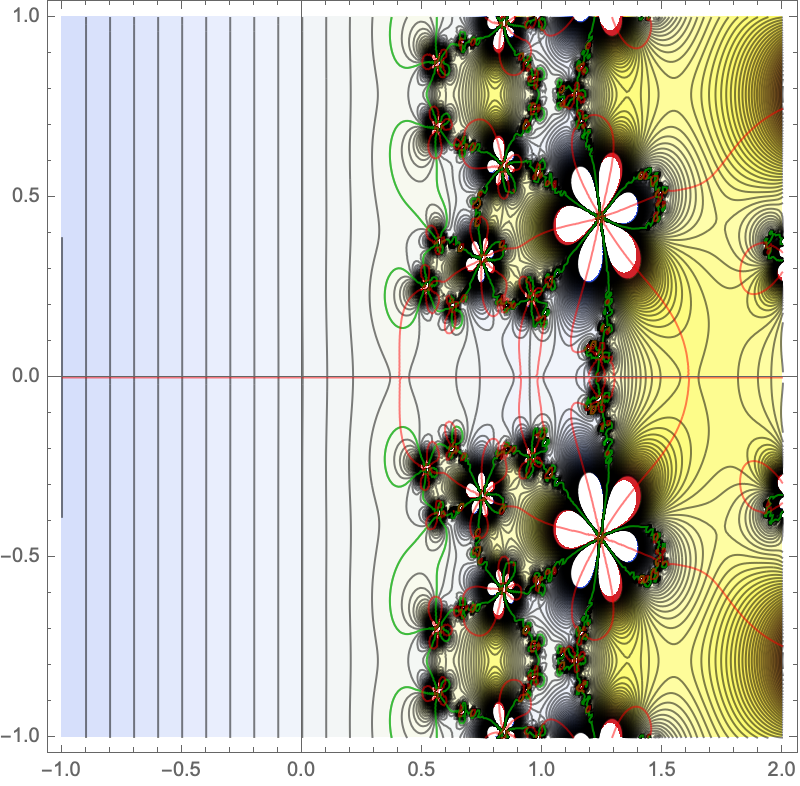}
        \caption{$n=5$}
    \end{subfigure}~
    \begin{subfigure}[t]{0.32\linewidth}
        \includegraphics[width =\textwidth]{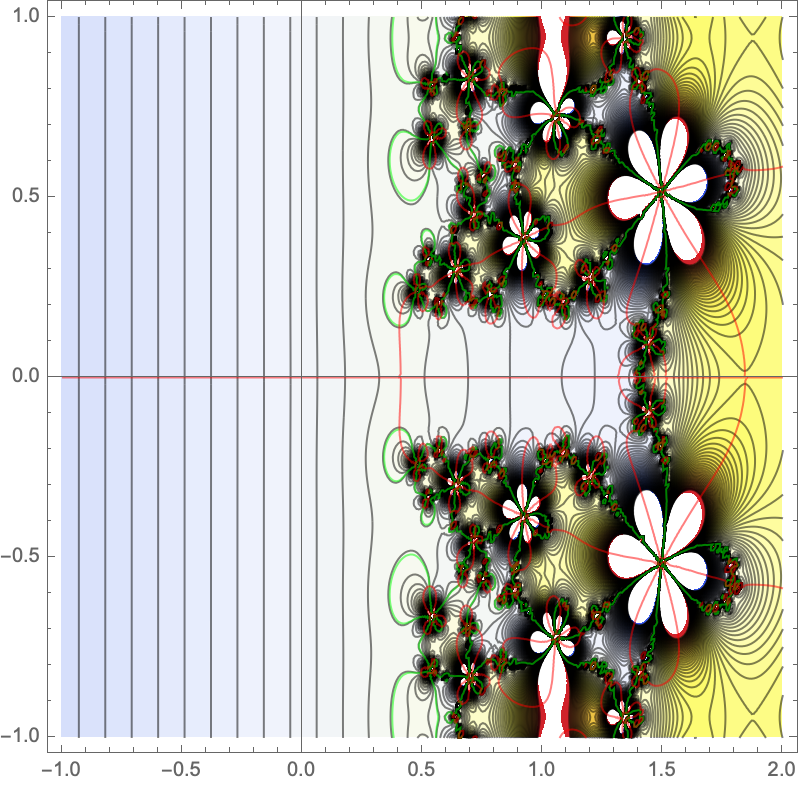}
        \caption{$n=6$}
    \end{subfigure}
    \caption{Comparison of the continuum and discretized classical paths. We plot the final position $x_1$ in the complex (continuum and discretized) initial velocity plane. We consider classical paths of a particle with mass $m=1$ starting at $x_0=-5$ propagating for a time $T=10$ while interacting with a Rosen-Morse barrier with strength $V_0=1$. The red curves correspond to the classical paths terminating at a real position. The orange curves correspond to their analytic continuations beyond the singularity crossings. The green lines represent the classical paths that terminate at a position whose real part is $-1$. The intersections of the red ($Im[x_1]=0$) and green lines ($Re[x_1]=-1$) are classical paths propagating between $x_0=-5$ and $x_1=-1$.}\label{fig:Discrete}
\end{figure*}

While the real solutions of the time-discretized theory 
\begin{align}
    \int e^{\frac{i \epsilon T}{\hbar}\sum_{j=0}^{n-1}\left[\frac{m}{2}\left(\frac{y_{j+1}-y_j}{\epsilon T}\right)^2 - V(y_j)\right] }\mathrm{d}y_1\mathrm{d}y_2\dots \mathrm{d}y_{n-1}\,,
\end{align}
approach the continuum theory 
\begin{align}
    G[x_1,x_0;T] = \int_{x(0)=x_0}^{x(1)=x_1} e^{\frac{i}{\hbar} \int_0^1 \left[ \frac{m \dot{x}^2}{2T} - T V(x) \right]\mathrm{d}\lambda}\mathcal{D}x
\end{align}
as the discretization is refined, the complex structure of the solutions is dramatically different (see fig.\ \ref{fig:Discrete}). 

In the continuum theory, the final position as a function of complex initial velocity develops a set of branch points and branch cuts (see fig.\ \ref{fig:continuumD}). Generally, there exists an infinite set of complex solutions to the boundary value problem roughly homogeneously distributed in the complex plane. We find that only a select few of these complex paths are relevant to the Feynman path integral \cite{Feldbrugge:2023Rosen}.

The final position in the discretized classical paths diverges for several complex discretized initial velocities $\bar{v}_0$. Note that the number of singularities increases as we increase the number of discretization steps $n$. Each singularity comes with a set of complex solutions to the boundary value problem, corresponding with the intersections of the red and green curves in fig.\ \ref{fig:Discrete}. It is an open problem which of these classical paths is relevant to the discretized path integral. One would hope that the complex classical path resembling the relevant continuous complex path dominates the discretized path integral.

\begin{figure}
    \includegraphics[width=\linewidth]{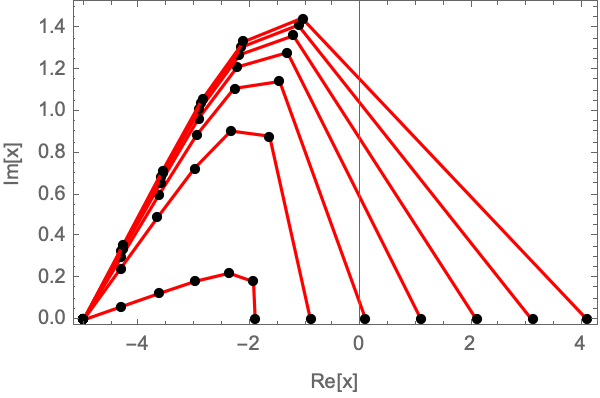}
    \caption{Discretized complex classical path for $n=6$ as the final position crosses past the singularity crossing of the continuum theory.}
    \label{fig:Discrete_Paths}
\end{figure}

However, this is probably not the full story, as the singularity crossings present in the continuum theory are absent in the discretized version. In particular, as we track the discretized classical path closest to the relevant continuum complex classical path while increasing the final position $x_1$, we find that the discretized classical path develops large jumps when the continuum classical path crosses the branch cut (see fig.\ \ref{fig:Discrete_Paths}). The path contains a gradual evolution followed by a large jump to the final position. These phenomena need to be studied in more detail in the future, as large jumps normally signal places where the discretization assumption is no longer valid. However, it could be that a good approximation of the continuum path integral is recovered with an additional set of discretized complex classical paths that are absent in the continuum theory.

\end{document}